\documentclass[aps,prl,twocolumn,groupedaddress]{revtex4}
\usepackage{graphicx}
\begin{document}

\title{Liquid elasticity length and universal dynamic crossovers}

\author{Kostya Trachenko$^{1}$}
\author{V. V. Brazhkin$^{2}$}
\address{$^1$ Department of Earth Sciences, University of Cambridge, UK}
\address{$^2$ Institute for High Pressure Physics, Russia}

\begin{abstract}
We discuss two main universal dynamic crossovers in a liquid that correspond to relaxation times of 1 ps and
$10^{-7}-10^{-6}$ s. We introduce the concept of liquid elasticity length $d_{el}$. At room temperature, $d_{el}$ is several
\AA\ in water and increases to 0.01 mm in honey and 1 mm in tar. We show that on temperature decrease, $d_{el}$ crosses the
fundamental lengths of the system, medium-range order $d_m$ and system size $L$. We discuss how $d_{el}=d_m$ and $d_{el}=L$
correspond to the two dynamic crossovers.
\end{abstract}


\maketitle

A conceptually simple phenomenon, freezing of liquid into glass, has turned out to be one of the most difficult problems in
condensed matter physics, the problem of glass transition \cite{langer}. Analyzing the current state of the field, Dyre
recently suggested that glass transition itself is not a big mystery: it universally takes in any liquid when its relaxation
time $\tau$ exceeds the time of experiment at the glass transition temperature $T_g$ \cite{dyre}. He proposed that the
challenges lie above $T_g$ \cite{dyre}.

If we consider the changes in molecular dynamics in a liquid on lowering the temperature, we find two dynamic crossovers.
The first crossover is at high temperature at $\tau\approx 1$ ps, at which dynamics changes from exponential Debye
relaxation to SER, $q(t)\propto \exp(-(t/\tau)^\beta)$, where $q$ is a relaxing quantity and $0<\beta<1$. This crossover is
universal, i.e. is seen in many systems \cite{exp1,exp2,exp3,exp4,exp5}. SER describes a very sluggish dynamics: in the wide
data range, it decays as a logarithm of time. As the temperature is lowered, we find another universal crossover, which
takes place at lower temperatures that correspond to $\tau=10^{-7}-10^{-6}$ s \cite{magic1,magic2}. This crossover also
marks the qualitative change in system's dynamics, and was attributed to the transition from the ``liquid-like'' to the
``solid-like'' behaviour. Note that although relaxation time at the second crossover is much larger as compared to the first
one, it is still about 9--10 orders magnitude smaller that relaxation time at the glass transition. By definition, $T_g$
corresponds to $\tau$ on the order of $10^3$ s.

One hopes to find useful insights into the problem of glass transition if the origin of dynamic crossovers could be
rationalized. In this paper, we discuss this problem in terms of elastic stresses that a liquid supports in response to an
external perturbation. We introduce the temperature-dependent liquid elasticity length, which is the range of elastic
interaction in a liquid. We propose that that the first and second crossovers take place when this length becomes equal to
the values of the medium-range length and system size, respectively.

We first note that in a perfect crystal, there are two fundamental lengths, lattice constant $a$ and system size $L$. In a
disordered system like liquid, there is an additional length $d_m$, which corresponds to the medium-range order, and is
defined by local packing. $d_m$ is on the order of 10 \AA, the characteristic size of decay of correlations in a disordered
media.

We now introduce the liquid elasticity length $d_{el}$. Unlike a solid, a liquid does not support static stresses. However,
it supports stresses at high frequencies in a solid-like manner \cite{dyre}. These frequencies correspond to times smaller
than system relaxation time $\tau$. If $d_{el}$ is the distance over which interactions in a liquid are elastic,
$d_{el}=c\tau$, where $c$ is the speed of propagation of elastic interactions, and is approximately equal to the speed of
sound. $c$ is on the order of $c=a/\tau_0$, where $\tau_0$ is the oscillation period, or inverse of Debye frequency
($\tau_0=0.1$ ps) and $a$ is the interatomic separation of about 1 \AA. We therefore find

\begin{equation}
d_{el}=\frac{\tau}{\tau_0}a
\end{equation}

We note that $d_{el}$ crosses, on lowering the temperature, all three fundamental lengths in a system. Since
$\tau=\tau_0\exp(V_0/kT)$ at high temperature, where $V_0$ is the high-temperature activation barrier,
$d_{el}=a\exp(V_0/kT)$. At high temperature, $d_{el}$ is on the order of interatomic distance $a$. On lowering the
temperature, it increases to $d_m$. Because $d_m$ is about 10 \AA, we find from Eq. (1) that $d_{el}=d_m$ gives $\tau$ of
about 1 ps. This is the first dynamic crossover discussed above. On lowering the temperature even further, Eq. (1) shows
that $d_{el}$ increases to $L$. In liquid relaxation experiments, $L$ is typically 1-10 mm. According to Eq. (1), $d_{el}=L$
gives $\tau=10^{-7}-10^{-6}$, the second dynamic crossover. This points to the intriguing possibility to describe the most
important changes in system's dynamics using only its three fundamental lengths.

To discuss why the two dynamic crossovers correspond to $d_{el}=d_m$ and $d_{el}=L$, we first recall the previous discussion
how a liquid relaxes stress in response to an external perturbation. Some time ago, Orowan introduced ``condordant'' local
rearrangement events \cite{orowan}. A concordant local rearrangement is accompanied by a strain agreeing in direction with
the applied external stress, and reduces the energy and local stress (see Figure 1). A discordant rearrangement, on the
other hand, increases the energy and local stress. This has led to a general result that stress relaxation by earlier
concordant events leads to the increase of stress on later relaxing regions in a system \cite{orowan}. Goldstein applied the
same argument to a viscous liquid \cite{gold}: consider a system under external stress which is counterbalanced by stresses
supported by local regions. When a local rearrangement to a potential minimum, biased by the external stress, occurs (a
concordant event), this local region supports less stress after the event than before; therefore, other local regions in the
system should support more stress after that event than before \cite{gold}.

\begin{figure}
\begin{center}
{\scalebox{0.7}{\includegraphics{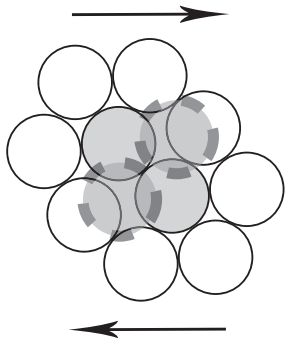}}}
\end{center}
\caption{Orowan's illustration of a concordant local rearrangement. Solid and dashed lines around the shaded atoms
correspond to initial and final positions of a rearrangement, respectively. Arrows show the direction of external stress.}
\end{figure}

Lets consider stress relaxation in a liquid under external perturbation (e.g., pressure or temperature). To illustrate the
argument of Orowan and Goldstein above, we consider a system under stress and discuss the dynamics of pressure-induced local
relaxation events (LREs), i.e those events that come in addition to thermal rearrangements. In a simple model of parallel
springs under external stress, a relaxation by a concordant event can be modeled as an appearance of a spring with a reduced
spring constant. It is easy to show \cite{recent1}, that after the introduction of $n$ weaker springs, which corresponds to
$n$ concordant relaxation events in a real system, the increase of stress on a current unrelaxed spring $\Delta p$ is
proportional to the current number of events $n$, $\Delta p\propto n$. We now invoke the previous discussion about the
activation barrier $V$ for a LRE in a real liquid. It has been argued that $V$ is essentially the elastic energy
\cite{dyre,nemilov,granato,dyre1}. In the Shoving model, for example, $V$ is given by the work of the elastic force needed
to shove aside the surrounding liquid in order for a relaxation event to take place \cite{dyre1}. Therefore the increase of
activation barrier is proportional to the increase of elastic force due to the increase of pressure, $\Delta p$, on a
current local rearranging region, and we find $V\propto\Delta p\propto n$. This provides the feed-forward interaction
mechanism between LREs, in that activation barriers increase for later events. This mechanism is general, and does not
depend on the choice of the spring model which we used here for illustration purposes only.

It is important to discuss the condition under which the feed-forward interaction mechanism becomes operative on lowering
the temperature. Let $t_{\rm s}$ be the time needed for elastic interaction to propagate between LREs, $d$ the distance
between LREs, $c$ the speed of sound, and $\tau_0$ the oscillation period, or inverse of Debye frequency ($\tau_0=0.1$ ps).
Because $c=a/\tau_0$, $t_{\rm s}=\tau_0 d/a$. Relaxation time $\tau$, which is also the time between two consecutive
relaxation events \cite{dyre}, increases on lowering the temperature. It is easy to see that at high enough temperature,
when $\tau=\tau_0$, $t_{\rm s}>\tau$ is always true because $d/a>1$. In this case, local events relax independently of each
other, because at high temperature, the time between the events is shorter than the time needed for elastic interaction to
propagate between them. Because events are independent, we obtain the expected high-temperature result that relaxation is
Debye (exponential) in time \cite{recent1} and Arrhenius in temperature. On cooling the system down, a certain temperature
always gives the opposite condition: $t_{\rm s}\le \tau$. When the time between local relaxation events exceeds the time of
propagation of elastic interaction between the events, the discussed above feed-forward interaction mechanism becomes
operative. In other words, when $t_{\rm s}\le\tau$, local relaxation events do not relax independently, but ``feel'' the
presence of each other. We have recently showed how this picture gives rise to SER \cite{recent1}. Note that $d_{el}$ sets
the maximal distance between those LREs which are involved in the feed-forward interaction mechanism: $t_{\rm s}=\tau$ gives
$d=a\tau/\tau_0=d_{el}$.

We are now ready to discuss how the first dynamic crossover appears in this picture. Lets denote $\tau_1$ the time of this
crossover. As follows from the above discussion, $\tau_1$ is equal to the time needed for elastic interaction to propagate
between neighbouring LRE. Let $d$ be the distance between two adjacent LRE. Its value is on the order of 10 interatomic
distances, or medium range order $d_m$. Because  $\tau_{1}=d_m/c$ and $c=a/\tau_0$, we find

\begin{equation}
\tau_{1}=\frac{d_m}{a}\tau_0
\end{equation}

We observe that $\tau_1$ is the relaxation time of the system which corresponds to $d_{el}=d_m$ (see Eq. (1)).

Because $\tau_0=0.1$ ps and $d_m/a$ are roughly system- and temperature-independent, Eq. (2) predicts that $\tau_{1}$ is a
universal parameter, independent on temperature or system type. This is consistent with experimental findings
\cite{exp1,exp2,exp3,exp4,exp5}. If $d/a$ is on the order of 10 and $\tau_0=0.1$ ps, we find from Eq. (2) that $\tau_1$ is
about 1 ps, in good agreement with the experimental value of $\tau_1$ in 1--2 ps range \cite{exp1,exp2,exp3,exp4,exp5}.

To discuss the second dynamic crossover, we elaborate on how, on lowering the temperature, the increase of the radius of the
feed-forward interaction increases activation barriers for LRE. As discussed above, the maximal distance between those LREs
which are involved in the feed-forward interaction mechanism is equal to the elasticity length $d_{el}$. According to Eq.
(1), $d_{el}$ is on the order of interatomic distances at high temperature, but increases on lowering the temperature. The
qualitative argument, which is sufficient for this discussion, proceeds as follows, and is illustrated in Figure 2. Let $N$
be the number of LREs, induced by external perturbation (i.e. those in addition to thermally-induced events) in the sphere
of diameter $d_{el}$. As discussed above, remote concordant LREs result in the increase of stress, $\Delta p$, on a current
local region. Because this increase is additive, $\Delta p$ increases with $N$. Since $N$ increases with $d_{el}$, $\Delta
p$ is a monotonously increasing function of $d_{el}$. As discussed above, the activation barrier $V\propto\Delta p$. Hence
we find that $V$ is a monotonously increasing function of $d_{el}$.

\begin{figure}
\begin{center}
{\scalebox{0.45}{\includegraphics{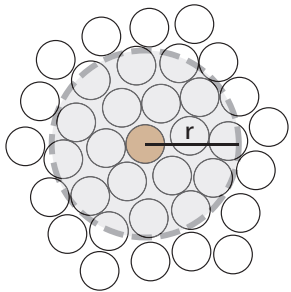}}}
\end{center}
\caption{Illustration of the feed-forward interaction mechanism between local relaxation events. This mechanism operates
within the sphere of radius $r=d_{el}/2$, which increases according to Eqs. (1-2). Shaded and open circles represent local
relaxing regions inside and outside, respectively, of the interaction sphere.}
\end{figure}

The second dynamic crossover originates in this picture when, on lowering the temperature, $d_{el}$ reaches the system size
$L$. At high temperature, only small parts of the system are elastic, $d_{el}\ll L$, as in the case of the first crossover
when $d_{el}=d_m$. When $d_{el}=L$ on lowering the temperature, all LREs in the system are involved in the feed-forward
interaction. Hence $d_{el}=L$ marks the transition of the system from being partially to wholly elastic, and should manifest
itself as a qualitative change in the liquid's dynamics.

Another interesting prediction from this picture comes from the observation that at $d>L$, reducing the temperature further
has a weak effect on $V$, because at this point the increase of $V$ is related to the feed-forward interaction between
temperature-induced events only, but not due to the increase $V$ by way of the increase of $d_{el}$. As a result, the system
is expected to show a crossover to more strong behaviour. Note that in this picture, the second crossover depends on system
size. If, for example, the increase of $V$ is described by the Vogel-Fulcher-Tamman law, the divergence at a finite
temperature is avoided because as the system crosses over to more strong behaviour, the divergence is pushed to zero
temperature. Only the infinite system does not have the second dynamic crossover, and has the divergence at a finite
temperature.

Experimentally, there is ample evidence for the dynamic crossover in many systems. Most direct evidence comes from the sharp
kink in the dielectric function \cite{magic1}. The crossover to the lower slope of relaxation time, with the effect that
glass transition becomes retarded, is observed \cite{stickel}, in agreement with the prediction from our picture. Other
experiments include NMR relaxation data, which detect a similar dynamic crossover \cite{maekawa}, the crossover in the
relaxation of cage sizes in the positron annihilation experiments \cite{ngai} and changes of non-ergodicity parameter
\cite{adin}.

If $\tau_2$ is relaxation time at the second crossover, $d_{el}=L$ gives (see Eq. (1)):

\begin{equation}
\tau_2=\frac{L}{a}\tau_0
\end{equation}

Eq. (3) predicts that, similar to $\tau_1$, $\tau_2$ is a universal parameter, independent on temperature or system type.
This is consistent with experimental findings \cite{magic1,stickel,maekawa,ngai,adin,magic2}. Typical values of $L$ used in
the experiments are 1--10 mm, which is dictated mostly by the experimental setup. For example, smaller system sizes can be
associated with surface effects, while larger system sizes can involve temperature gradients. Using this range of $L$ and
$\tau_0=0.1$ ps, we find from Eq. (3) that $\tau_2=10^{-7}-10^{-6}$ s, in good agreement with experimental values
\cite{magic1,magic2}. Note that at the second $d_{el}=L$ crossover, $1/\tau_2$ has the meaning of the eigenfrequency of the
system.

We note that relaxation time $\tau_2=10^{-7}-10^{-6}$ s corresponds to viscosities $\eta$ of $10^3-10^4$ Pa$\cdot$s. These
are much larger than room-temperature viscosities of familiar liquids like water, ethanol, olive oil or glycerol, for which
$\eta=10^{-3}-10^1$ Pa$\cdot$s. For these liquids, Eq. (1) gives $d_{el}$ in the range of 1--1000 Angstroms. Viscosity of
honey ($\eta=10^2$ Pa$\cdot$s) still falls behind the viscosities that correspond to relaxation time $\tau_2$. Honey's
elasticity length is $d_{el}=0.01$ mm. Very viscous tar ($\eta=10^4$ Pa$\cdot$s, $d_{el}$=1 mm) has relaxation time close to
$\tau_2$ and elasticity length comparable to the experimental length scale. Only extremely viscous pitch ($\eta=10^8$
Pa$\cdot$s, $d_{el}$=10 m) has relaxation time that exceeds $\tau_2$ and elasticity length in excess of the experimental
length scale. Hence these examples illustrate that although the second crossover is way above the glass transition (see
below), it corresponds to quite high values of viscosity. They also give the feeling for the actual values of the introduced
elasticity length in real liquids.

In this picture, no dynamic crossover takes place at $T_g$, consistent with the experimental observations. According to Eq.
(1), $\tau(T_g)=10^3$ s corresponds to $d_{el}$ of 1000 kilometers, which explains why the two dynamic crossovers are seen
way before $T_g$ is reached. In other words, the absence of a dynamic crossover in the vicinity of $T_g$ is due to the
imbalance between our typical experimental times and sample sizes: at typical experimental time, the elasticity length is
more than 8 orders of magnitude larger than the typical experimental length. In this context, we note that thermodynamic
anomalies at $T_g$ are sensitive to cooling rates and observation times. Were the thermodynamic parameters to be measured at
higher cooling rates and shorter experimental times, they would show the changes at temperatures that correspond to
relaxation time $\tau_2$.

In summary, we have introduced liquid elasticity length, and showed that when it becomes equal to the medium-range length
and system size, two universal dynamic crossovers take place. The values of crossovers are fixed by $a$, $d_m$ and $L$,
three fundamental lengths in a disordered system. From the physical point of view, the first crossover can be said to be
more fundamental since it does not depend on system size $L$. Our finding represents a new interesting link between geometry
and dynamics of the system.

We are grateful to R. Casalini and M. Roland for discussions and to EPSRC for support.


\begin{thebibliography}{99}

\bibitem{langer} J. Langer, Physics Today, p. 8 (February 2007).

\bibitem{dyre} J. C. Dyre, Rev. Mod. Phys. {\bf 78}, 953 (2006).

\bibitem{exp1} C. M. Roland, K. L. Ngai and L. J. Lewis, J. Chem. Phys. {\bf 103}, 4632 (1995).

\bibitem{exp2} K. L. Ngai and G. N. Greaves, J. Non-Cryst. Sol. {\bf 182}, 172 (1995)

\bibitem{exp3}  J. Colmenero, A. Arbe and A. Alegria, Phys. Rev. Lett. {\bf 71}, 2603 (1993)

\bibitem{exp4} J. Colmenero et al, Phys. Rev. Lett. {\bf 78}, 1928 (1997)

\bibitem{exp5} R. Zorn et al, Phys. Rev. E {\bf 52}, 781 (1995).

\bibitem{magic1} A. Sch\H{o}nhals, Europhys. Lett. {\bf 56}, 815 (2001).

\bibitem{magic2} V. N. Novikov and A. P. Sokolov, Phys. Rev. E {\bf 67}, 031507 (2003).

\bibitem{orowan} E. Orowan, Proceedings of the First National Congress of Applied Mechanics (American Society of Mechanical Engineers, New York), 453 (1952).

\bibitem{gold} M. Goldstein, J. Chem. Phys. {\bf 51}, 3728 (1969).

\bibitem{recent1} K. Trachenko, arXiv:cond-mat/0611648v1.

\bibitem{nemilov} S. V. Nemilov, J. Non-Cryst. Sol. {\bf 352}, 2715 (2006).

\bibitem{granato} A. V. Granato and V. A. Khonik, Phys. Rev. Lett. {\bf 93}, 155502 (2004).

\bibitem{dyre1} J. C. Dyre, N. B. Olsen and T. Christensen, Phys. Rev. B {\bf 53}, 2171 (1996).

\bibitem{stickel} F. Stickel, E. W. Fischer and R. Richert, J. Chem. Phys. {\bf 104}, 2043 (1996).

\bibitem{maekawa} H. Maekawa, Y. Inagaki, S. Shimokawa and T. Yokokawa, J. Chem. Phys. {\bf 103}, 371 (1995).

\bibitem{ngai} K. L. Ngai, L. R. Bao, A. F. Lee and C. L. Soles, Phys. Rev. Lett. {\bf 87}, 215901 (2001).

\bibitem{adin} S. V. Adichtchev et al, Phys. Rev. Lett. {\bf 87}, 055703 (2001).

\end{thebibliography}
\end{document}